\documentclass{elsart}
\usepackage{graphicx} 
\begin{document}
\begin{frontmatter} 

\title{Roughness fluctuations, roughness exponents and the universality class
of ballistic deposition}
\author{F. D. A. Aar\~ao Reis}
\ead{reis@if.uff.br}           
\address{
Instituto de F\'\i sica, Universidade Federal Fluminense,\\
Avenida Litor\^anea s/n, 24210-340 Niter\'oi RJ, Brazil}
\date{\today}
\maketitle

\begin{abstract}
In order to estimate roughness exponents of interface growth models, we
propose the calculation of effective exponents from the roughness
fluctuation $\sigma$ in the steady state. We compare the finite-size behavior
of these exponents and the ones calculated from the average
roughness $\langle w_2\rangle$ for two models in the $2+1$-dimensional
Kardar-Parisi-Zhang (KPZ) class and for a model in the $1+1$-dimensional
Villain-Lai-Das Sarma (VLDS) class. The values obtained from
$\sigma$ provide consistent asymptotic estimates, eventually with smaller
finite-size corrections. For the VLDS (nonlinear molecular beam epitaxy)
class, we obtain $\alpha=0.93\pm 0.01$, improving previous estimates. We
also apply this method to two versions of the ballistic deposition model in
two-dimensional substrates, in order to clarify the controversy on its
universality class raised by numerical results and a recent derivation of its
continuous equation. Effective exponents calculated from $\sigma$ suggest that
both versions are in the KPZ class. Additional support to this conclusion is
obtained by a comparison of the full roughness distributions of those models
and the distribution of other discrete KPZ models.
\end{abstract}


\end{frontmatter}

\section{Introduction}
\label{intro}

Surface and interface growth processes attract much interest 
for their applications in solid state physics and material
science~\cite{frontiers,barabasi,krug}. They motivated the study of
many interface growth models, which led to significant advances in
non-equilibrium Statistical Mechanics and related fields~\cite{barabasi,krug}. 

The simplest quantitative characteristic of a given interface is its
roughness, also called the interface width, defined as the rms fluctuation of
the height around its average position. The squared width,
$w_2\equiv \overline{h^2} - {\overline{h}}^2$, is usually averaged over
different configurations, and its scaling on time and length is used to
characterize the growth process. For short times, in the so-called growth
regime, the average roughness increases as ${\langle w_2\rangle}_{gr} \sim
t^{2\beta}$, where $\beta$ is the growth exponent.
For long times in a finite substrate, a steady state is attained, where
the average roughness scales with the linear size $L$ as
\begin{equation}
\langle w_2\rangle \sim L^{2\alpha} ,
\label{defalpha}
\end{equation}
with $\alpha$ called the roughness exponent (notice that steady
state averages are denoted without further labels here, while indexes $s$ or
$sat$ are frequently used in the literature to denote saturation). The
connection of discrete or continuous growth models to certain growth equations
is frequently based on accurate numerical calculation of those scaling
exponents. However, this procedure may be quite difficult when huge
finite-size or finite-time corrections are present in the scaling of $\langle
w_2\rangle$.

Alternatively, one may investigate features of the
full roughness distribution, $P\left( w_2\right)$, which is the probability
density of the roughness of a given configuration to lie in the range $\left[
w_2, w_2+dw_2\right]$. In recent years, the steady state distributions of
several interface models have been
analyzed~\cite{foltin,racz,antal,distrib} and they were shown
to fit the scaling form 
\begin{equation}
P\left( w_2\right) = {1\over
\langle w_2\rangle} \Phi\left( {w_2\over{\langle w_2\rangle}}\right) .
\label{scaling1}
\end{equation}
Alternatively, the scaling relation
\begin{equation}
P\left( w_2\right) =
{1\over \sigma} \Psi\left( {{w_2-\left< w_2\right>}\over\sigma}\right)
\label{scaling2}
\end{equation}
can be adopted, where
\begin{equation}
\sigma \equiv \sqrt{ \left< {w_2}^2 \right> - {\left< w_2\right>}^2 }
\label{defsigma}
\end{equation}
is the rms deviation of the squared roughness. Despite the study of
roughness distributions being useful for determining the universality classes
of some growth processes, it is usually performed independently of the
calculation of scaling exponents (notice that the values of the exponents are
not necessary to fit the scaled distributions). Indeed,
finite-size corrections in the universal functions $\Phi$
and $\Psi$ are typically much smaller than those in the
Family-Vicsek scaling relation~\cite{fv}, which gives Eq.
(\ref{defalpha}) a limiting case.

On the other hand, from Eqs. (\ref{scaling2}) and (\ref{defsigma}), we expect
that the roughness fluctuation $\sigma$ in the steady state scales with
exponent $\alpha$ as the average roughness:
\begin{equation}
\sigma\sim L^{2\alpha} .
\label{scalingsigma}
\end{equation}
Consequently,
numerical estimates of $\sigma$ may also be used to estimate $\alpha$, but, as
far as we know, this was not done in previous works. The expectation of weaker
finite-size corrections using $\sigma$ instead of $\left< w_2\right>$ follows
from the recent observation~\cite{distrib} that Eq. (\ref{scaling2}) provides
a better data collapse of simulation data when compared to Eq. (\ref{scaling1}),
for several growth models.

In order to estimate a scaling exponent with accuracy, the standard method is
to extrapolate effective exponents obtained from consecutive slopes of
log-log plots (see, e. g., Appendix A of Ref. \protect\cite{barabasi}). Here,
we will show that the effective exponents obtained from $\sigma$ provide
reliable estimates of the roughness exponent and, in some cases, varies
with $L$ slower than the estimates from the
average width $\left< w_2\right>$. This is illustrated with applications to
three growth models which belong to different universality classes, in one-
and two-dimensional substrates.

Subsequently, we will apply this method to calculate the roughness
exponent of two versions of the ballistic depositon (BD) model~\cite{vold} in
two-dimensional substrates. Numerical works showed discrepancies in the
estimates of that exponent in $d=2$ and a rigorous derivation of a continuous
equation for BD~\cite{katzav} deviates from the Kardar-Parisi-Zhang
(KPZ) equation~\cite{kpz}. However, when effective exponents of BD
are calculated with $\sigma$, the asymptotic estimates of $\alpha$ are close to
the best known KPZ values. The conclusion that this is the universality class
of BD is reinforced by the comparison of the full roughness distributions.

The rest of this work is organized as follows. In Sec. 2 we present the
models which will be used to test the method for estimating roughness
exponents and the results of these tests. In Sec. 3 we apply that method to
two versions of BD in $d=2$ and analyze their roughness distributions. In Sec.
4 we summarize our results and conclusions.

\section{Tests of the method in models of the KPZ and the VLDS classes}
\label{secmodel}

The first model analyzed here is the RSOS model~\cite{kk}, in which the 
incident particle can stick at the top of a column only if the differences of
heights of all pairs of neighboring columns do not exceed ${\Delta H}_{max} =
1$ after aggregation. Otherwise, the aggregation attempt is rejected. The
second model was proposed for etching of a crystalline solid by Mello
et al~\cite{mello}: at each growth attempt, a randomly chosen column $i$, with
current height $h(i)\equiv h_0$, has its height increased of one unit
($h(i)\rightarrow h_0+1$), and all neighboring column whose heights are smaller
than $h_0$ grows to $h_0$ (this may be called the growth version of the
etching model~\cite{kpz2d}).

In the limit of large lengths and long times, these models are represented by
the KPZ equation~\cite{kpz}
\begin{equation}
{{\partial h}\over{\partial t}} = \nu_2{\nabla}^2 h + \lambda_2
{\left( \nabla h\right) }^2 + \eta (\vec{x},t) .
\label{kpz}
\end{equation}
Here, $\nu_2$ and $\lambda_2$ are constants and $\eta$ is a Gaussian noise with
zero mean and variance $\langle \eta\left(\vec{x},t\right)
\eta\left(\vec{x'},t'\right) \rangle = D\delta^d \left(\vec{x}-\vec{x'}\right)
\delta\left( t-t'\right)$, where $d$ is the dimension of the substrate. In
$d=1$, the exact roughness exponent is $\alpha=1/2$~\cite{kpz,barabasi}, and
in $d=2$ the best numerical estimates are around
$0.38-0.39$~\cite{marinari,kpz2d}. Numerical estimates of $\left< w_2\right>$
for the RSOS and etching models in $d=1$ provide estimates of
exponents which converge to the exact values with negligible
finite-size corrections. However, that is not the case in $d=2$, particularly
for the etching model~\cite{kpz2d}, consequently our tests will focus this
case.

The widths $\left< w_2\right>$ and their fluctuations $\sigma$ were
calculated at the steady states in two-dimensional substrates up to $L=256$
for the RSOS model and up to $L=512$ for the etching model.
Finite-size estimates of the roughness exponents are given by
\begin{equation}
\alpha_w\left( L\right) \equiv {1\over 2} {
\ln{ \left[{\langle w_2\rangle} \left( L\right)
/ {\langle w_2\rangle}\left( L/2\right)\right] }\over \ln{2} }
\label{defalphaw}
\end{equation}
and
\begin{equation}
\alpha_\sigma\left( L\right) \equiv {1\over 2} {
\ln{ \left[\sigma \left( L\right)
/ \sigma\left( L/2\right)\right] }\over \ln{2} } .
\label{defalphas}
\end{equation}
The presence of significant corrections to Eqs. (\ref{defalpha})
and (\ref{scalingsigma}) is reflected in a size-dependence of $\alpha_w\left(
L\right)$ and $\alpha_\sigma\left( L\right)$.
These effective exponents are shown in Fig. 1 for the RSOS and the etching
models as a function of $1/L$ and $2/L$, respectively. 

An asymptotic estimate near $0.39$ is consistent with the trend of the RSOS
data as $L\to\infty$. For this model, small differences in the estimates
obtained from $\left< w_2\right>$ and $\sigma$ are observed, as well as weak
finite-size effects. The linear fit of $\alpha_w\left( L\right)$
shown in Fig. 1 suggests $\alpha\approx 0.39$.

However, the finite-size dependence of the data for the etching model is
remarkable: the effective exponents vary between $0.34$ and $0.38$ in Fig. 1.
For the largest lengths $L$, $\alpha_\sigma\left( L\right)$
approaches the asymptotic region $\alpha\approx 0.39$ faster than
$\alpha_w\left( L\right)$, as illustrated by the curves through the data
points in Fig. 1. It suggests that, for large $L$, the corrections to
the scaling of $\sigma$ are smaller than the corrections to 
$\left< w_2\right>$. Although the fits of the data in Fig. 1 do not represent
systematic extrapolation procedures, our interpretation is reinforced by
the fact that the extrapolated $\alpha_\sigma\left( L\right)$ for the etching
model is very close to the value predicted by the RSOS model.

\begin{figure}[ht]
\centering
\includegraphics[clip,width=0.7\textwidth, 
height=0.42\textheight,angle=0]{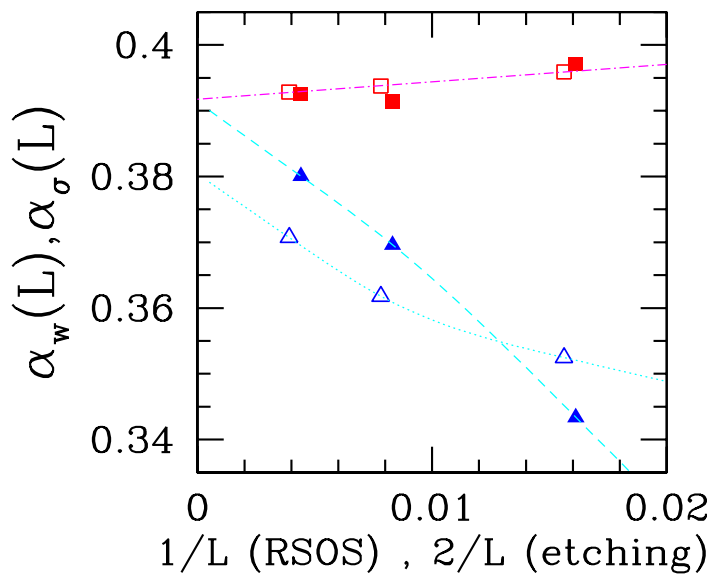}
\caption{Effective exponents $\alpha_w\left( L\right)$ (empty symbols) and
$\alpha_\sigma\left( L\right)$ (full symbols), for the RSOS model (squares) and
the etching model (triangles) in two-dimensional substrates, as a function of
$1/L$ or $2/L$. $\alpha_\sigma\left( L\right)$ data are slightly shifted to
the right to avoid superposition of data points. Error bars are compared to
the size of the points, except for the largest lengths,
in which the uncertainties are around $0.005$ for both models.
The lines connecting the data points (except $\alpha_\sigma\left( L\right)$
data for the RSOS model) were drawn to guide the eye.}
\label{fig1}
\end{figure}     

For the etching model, we were not able to
perform reliable extrapolations of
$\alpha_w\left( L\right)$ or $\alpha_\sigma\left( L\right)$ as functions of
$1/L^{\Delta}$, with any exponent $\Delta >0$ (the value $\Delta
=1$ is used in Fig. 1). This is related to the large deviations from the
asymptotic scaling in small lattices, typically $L<50$. It contrasts to
Ref. \protect\cite{kpz2d}, where a extrapolation of $\alpha_w\left(
L\right)$ up to $L=1024$ was possible with $\Delta\sim 0.5$. However,
here we are working with data for $L\leq 256$ because there are very
large error bars in the estimates of $\sigma$ for larger lattices. With a
small number of data points, even more sophisticated extrapolation
methods~\cite{henkel} are not able to provide better estimates than previous
work.

An extended version of the conserved RSOS (CRSOS) model is also suitable for
this test. In this model, if the aggregation at the column of incidence
disobeys the condition ${\Delta H}_{max}=1$, then the incident particle
executes a random walk among neighboring columns until finding a column in
which it can aggregate~\cite{reiscrsos}. The CRSOS model belongs to the class
of the nonlinear molecular-beam epitaxy equation, or  Villain-Lai-Das Sarma
(VLDS) equation~\cite{villain,laidassarma}
\begin{equation}
{{\partial h}\over{\partial t}} = \nu_4{\nabla}^4 h +
\lambda_{4} {\nabla}^2 {\left( \nabla h\right) }^2 + \eta (\vec{x},t) ,
\label{vlds}
\end{equation}
where $\nu_4$ and $\lambda_{4}$ are constants. Our extended version of the
CRSOS model is different from the original one~\cite{crsos1} but has the same
symmetries and, consequently, belong to the same universality class.

In Fig. 2 we show the effective
exponents of this model obtained from simulations up to $L=256$, with linear
fits of both data sets.  The evolution of $\alpha_\sigma\left( L\right)$ and
$\alpha_w\left( L\right)$ suggest $\alpha\approx 0.93$ asymptotically, but the
difference between $\alpha_\sigma\left( L\right)$ and the asymptotic value is
clearly much smaller.  This is consistent with the different slopes of the
linear fits shown in Fig. 2 (smaller slope for fitting $\alpha_\sigma\left(
L\right)$).

\begin{figure}[ht]
\centering
\includegraphics[clip,width=0.8\textwidth, 
height=0.48\textheight,angle=0]{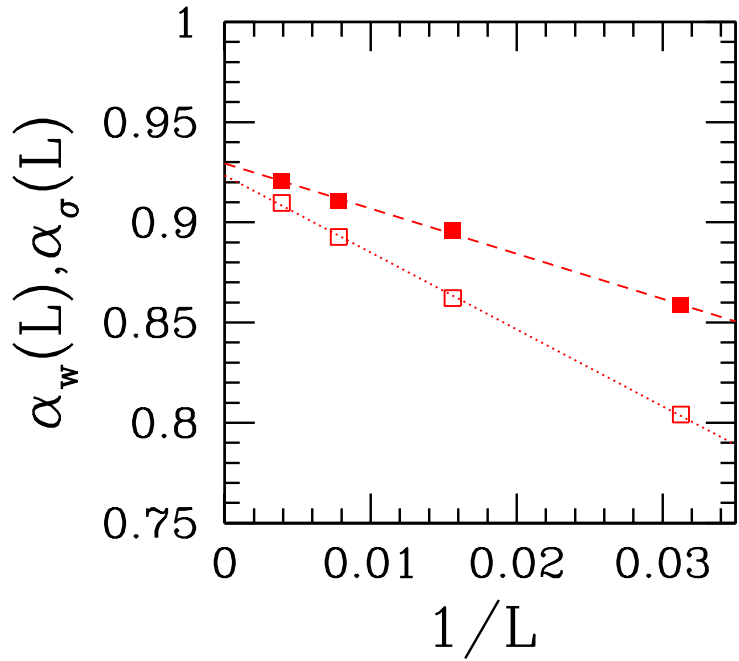}
\caption{Effective exponents $\alpha_w\left( L\right)$ (empty squares) and
$\alpha_\sigma\left( L\right)$ (full squares), for the CRSOS model in
one-dimensional substrates, as a function of $1/L$. Error bars are of the size
of the data points. The dotted and dashed lines are linear fits of each data
set.}
\label{fig2}
\end{figure}     

This application provides additional support to the conclusion that $\alpha
<1$ in the VLDS class in $d=1$~\cite{janssen} ($\alpha=1$ is obtained from
one-loop renormalization). The estimate $\alpha=0.93\pm 0.01$ follows from the
extrapolation of those data. It improves our
previous estimate $\alpha=0.94\pm 0.02$, obtained in Ref.
\protect\cite{reiscrsos}, where effective exponents were calculated from the
second and fourth moments of the height distribution in sizes up to $L=1024$.

The above analysis, performed with models in different universality classes
and with noticeable (but not huge) finite-size corrections to Eq.
(\ref{defalpha}), shows that roughness exponents may be estimated from
$\sigma$ with comparable or higher accuracy than those obtained from
$w_2$, suggesting its application to controversial situations.

\section{Application to ballistic deposition}
\label{ballistic}
  
In the simplest version of the ballistic deposition model, particles are
released from a randomly chosen position above a $d$-dimensional substrate,
follow trajectories perpendicular to the surface and stick upon first
contact with a nearest neighbor occupied site~\cite{fv,vold}. The
resulting aggregate is porous and has a rough surface. Several applications of
the BD model or its extensions to real growth processes were already proposed,
which justifies the present analysis (see e. g. recent applications in
Refs. \protect\cite{trojan,grzegorczyk}).

The mechanism of lateral aggregation suggests that BD is in the KPZ
class. However, several numerical works on this model showed discrepancies
between the estimated exponents and the KPZ
values~\cite{fv,ball,raissa,miranda}. In a recent work, we have shown that the
KPZ values are obtained with good accuracy in $d=1$ if effective exponents
$\alpha_w\left( L\right)$ are properly extrapolated~\cite{balfdaar}. 
In $d=2$, the effective roughness exponents rapidly vary with
$L$, but they are still far from the region $[0.38,0.40]$
up to $L=512$~\cite{balfdaar}. An apparent consistency with the KPZ value is
obtained in $d=2$ only after assuming of a constant correction term in Eq.
(\ref{defalpha}) (the intrinsic width), but error bars are very
large~\cite{kpz2d}. Consequently, from numerical work, there is
no clear evidence that BD is in the KPZ class in $d=2$.

Recently, Katzav and Schwartz derived a continuous equation for another
version of BD which differs from the KPZ equation~\cite{katzav}. They analyzed
the BD model with next-nearest neighbor aggregation, hereafter called BDNNN, in
which the incident particle sticks upon first contact with a nearest neighbor
or with a next-nearest neighbor occupied site. In $d=1$, they showed that the
squared gradient in Eq. (\ref{kpz}) is replaced by the absolute value
of the gradient in the continuous equation corresponding to BDNNN. This is
suggested as a possible reason for the slow convergence of the scaling
exponents to their asymptotic values. In $d\geq2$ the situation is more
complicated: the only contribution to the local growth rate comes from the
direction $x_m$ along which the height gradient is maximum, and that
contribution is ${1\over 2}{{\partial^2 h}\over{\partial {x_m}^2}} + \left|
{{\partial h}\over{\partial x_m}}\right|$. This accounts for the maximal
growth ingredient of the model. However, the presence of the lattice axis in
that equation contrasts with the rotation symmetry of the KPZ equation. Thus,
also from analytical work, there is no a priori reason to expect BD to
be in the KPZ class~\cite{katzav}.

This scenario motivates the extension of our method to estimate roughness
exponents of BD and BDNNN in $d=2$. Simulations of both models were performed
in lattices up to $L=512$, where $w_2$ and $\sigma$ can be obtained with
reasonable accuracy. The number of different realizations used to estimate
those quantities were near ${10}^8$ for $L\leq 128$, ${10}^7$ for $L=256$ and
$2\times {10}^6$ for $L=512$.

In Fig. 3a we show $\alpha_w\left( L\right)$ and $\alpha_\sigma\left( L\right)$
for the original BD model. $\alpha_w\left( L\right)$ rapidly varies with
$L$ in the whole range of our data and attains a value near $0.28$ in the
largest lattice, which is much smaller than the estimates for the
KPZ class.
This complex finite-size behavior was already pointed out in Ref.
\protect\cite{balfdaar} and, consequently, no reliable extrapolation of
$\alpha$ could be extracted from those data. On the other hand,
$\alpha_\sigma\left( L\right)$ is in the range $[0.36,0.37]$ for the largest
$L$. As $1/L\to 0$, Fig. 3a indicates that
$\alpha_\sigma\left( L\right)$ converges to an asymptotic value
between $0.36$ and $0.4$.

In Fig. 3b we show the effective exponents for the BDNNN model. We notice
that $\alpha_w\left( L\right)$ is even farther from the KPZ estimate in this
model. However, the trend of $\alpha_\sigma\left( L\right)$ as $1/L\to 0$
suggests an asymptotic value between $0.35$ and $0.4$, although finite-size
effects seem to be much stronger than those of the original BD model.

\begin{figure}[ht]
\centering
\includegraphics[clip,width=1.0\textwidth, 
height=0.60\textheight,angle=0]{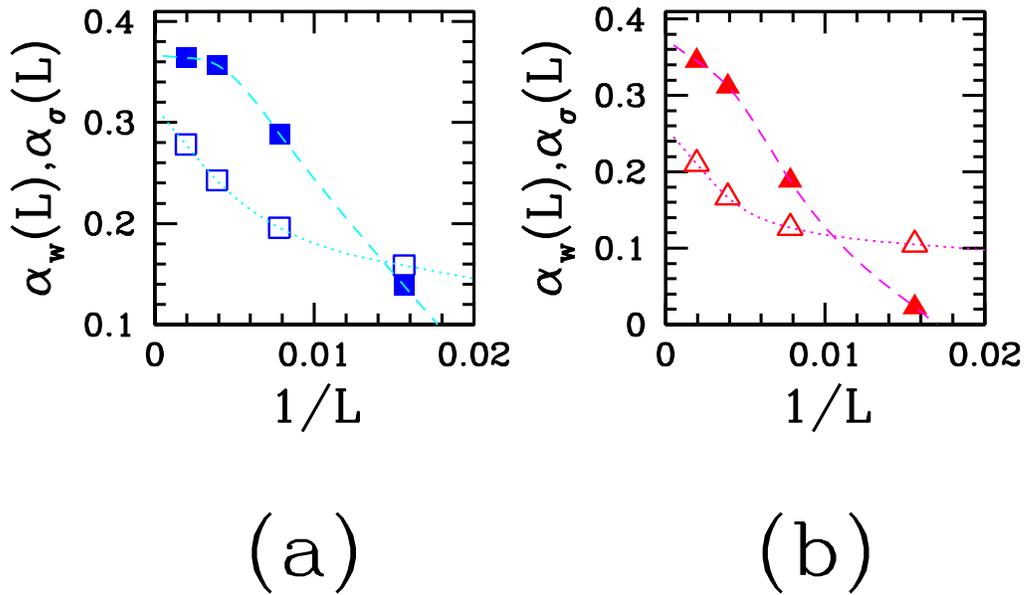}
\caption{Effective exponents $\alpha_w\left( L\right)$ (empty symbols) and
$\alpha_\sigma\left( L\right)$ (full symbols) for: (a) the original BD model in
$d=2$; (b) the BDNNN model in $d=2$. The lines connecting the data points were
drawn to guide the eye.}
\label{fig3}      
\end{figure}     

Extrapolations of $\alpha_w\left( L\right)$ and $\alpha_\sigma\left(
L\right)$ for BD and BDNNN as functions of $1/L^\Delta$, varying the exponent
$\Delta$ along the same lines of Ref. \protect\cite{kpz2d}, were also tested,
but no significant improvement of the
above mentioned estimates was obtained.
However, those estimates are consistent
with the best currently avaiable values for the KPZ class, which lie in the
range $\left[ 0.375,0.396\right]$~\cite{marinari,kpz2d}, which strongly
suggests that BD and BDNNN are also in that class in $d=2$.

Another important test is a comparison of the full roughness distributions of
these models and that of other KPZ models. The distribution for the RSOS
model, which shows negligible finite-size effects, can be used as a
representative of the KPZ class~\cite{distrib}. In Fig. 4 we
show this distribution, scaled according to Eq. (\ref{scaling2}), and the
scaled distributions for the BD and the BDNNN models in $L=256$ (data for
$L=512$ was not used here due to their lower accuracy, mainly in the tails of
the distribution). The collapse of all data into a single universal curve
reinforces our conclusion that both models are in the KPZ class in $d=2$.

\begin{figure}[ht]
\centering
\includegraphics[clip,width=0.8\textwidth, 
height=0.48\textheight,angle=0]{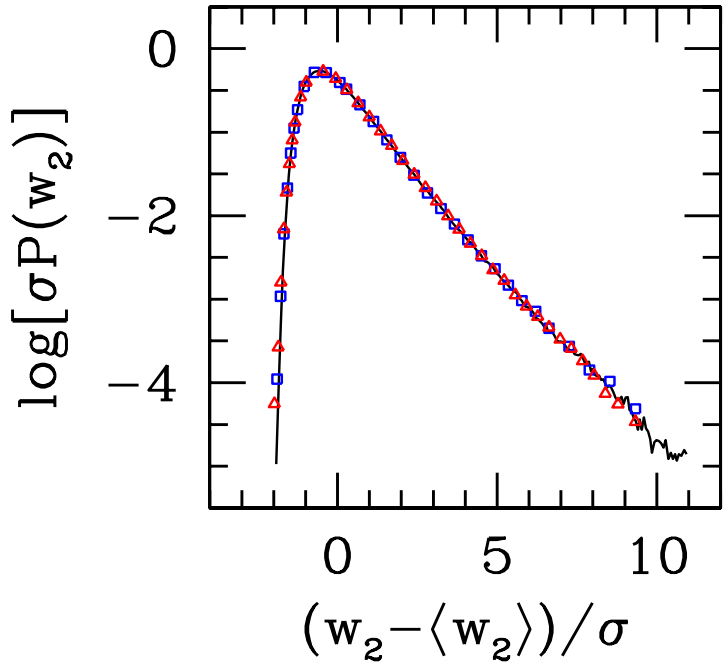}
\caption{Scaled roughness distribution at the steady states of the RSOS model
(solid curve), the BD model (squares) and the BDNNN model (triangles), in
$2+1$-dimensional lattices of length $L=256$.}
\label{fig4}
\end{figure}

A quantitative support to the universality of the scaled distributions of Fig.
4 is given by the estimates of the skewness and kurtosis of those curves. For
the RSOS and the etching models, $S= 1.70\pm 0.02$ and $Q=5.4\pm 0.3$ were
obtained in Ref. \protect\cite{distrib}. Here, for BD we obtained $S=1.71\pm
0.02$ and $Q=5.4\pm 0.1$, and for BDNNN we obtained $S=1.67\pm 0.03$ and
$Q=5.2\pm 0.3$, both in good agreement with the values of the other KPZ models.

\section{Conclusion}
\label{conclusion}

We proposed the calculation of roughness exponents from the scaling of
roughness fluctuations $\sigma$ in the steady states. Tests for two models in
the $2+1$-dimensional KPZ class and for a model in the $1+1$-dimensional VLDS
class were performed and showed that effective exponents obtained from that
quantity provide comparable or better asymptotic estimates than the
average width $\langle w_2\rangle$. We also applied this method to two versions
of the ballistic deposition model, the original one (BD) and the next-nearest
neighbor one (BDNNN), in order to clarify a controversy on its universality
class raised by numerical results and an exact derivation of its
corresponding continuous equation. Effective exponents calculated from $\sigma$
suggest that the asymptotic class is KPZ, while those calculated from $\langle
w_2\rangle$ show much more complex finite-size behavior. Additional support to
this conclusion is obtained from the collapse of the full roughness
distributions of those models into the distribution of the RSOS model.

At this point it is important to recall that higher moments of the
height distribution, such as the fourth moment $\left< w_4\right> \equiv
\left<\overline{{h - \overline{h}}^4}\right>$, are also frequently used to
estimate the roughness exponent. Their accuracy is usually lower
than that of $\left< w_2\right>$ and, consequently, do not improve the
estimates of $\alpha$ (see e. g. Ref. \protect\cite{kpz2d}). This quantity,
however, must not be confused with $\sigma^2$, which is the fluctuation of the
global roughness in the steady state.

The successful application of the above method to estimate the roughness
exponent, possibly combined with the analysis of roughness distributions,
suggests its extension to other models in which remarkable finite-size
corrections in the effective exponents are observed.


\end{document}